\documentstyle[12pt]{report}
\begin{document}
\def\hh{\hrule height0.9pt width1.1em}
\def\vv{\vrule width0.8pt depth0.12em height0.92em}
\def\square{vbox{\kern0.15em\hh\kern0.9em\hh\kern -1.05em\hbox{\vv\kern0.93em
\vv}}}
\def\frame#1#2#3#4{\vbox{hrule height#1pt%
\hbox{\vrule width#1pt\kern#2pt%
\hbox{\kern#2pt%
\vbox{\hsize#3\noindent#4}%
\kern #2pt}%
\kern#2pt\vrule width 1pt}%
\hrule\height0pt depth#1pt}}%
\def\s#1{\frame{.3}{.5}{5pt}{\centerline{#1\vphantom{(} } } )}
\def\mathbf{\vec}

\centerline{\Large {\bf Anomalous Pion Decay Revisited}}

\vskip0.3cm

\centerline{A.O. Battistel*, G. Dallabona*, B. Hiller**, M.C. Nemes*}

\vskip1cm

\centerline{* Departamento de F\'\i sica} 

\centerline{Universidade Federal de Minas Gerais} 

\centerline{CP 702, CEP 30161-970, Belo Horizonte, MG, Brazil}

\vskip1cm

\centerline{** Universidade de Coimbra}

\centerline{Centro de F\'\i sica Te\'orica} 

\centerline{3000 Coimbra, Portugal}

\vskip1cm

\vskip3cm

\centerline{\bf Abstract}

\vskip0.4cm

An implicit four dimensional regularization is applied to calculate 
the axial-vector-vector anomalous amplitude. The present technique always 
complies with results of Dimensional Regularization and can be easily applied
to processes involving odd numbers of $\gamma_5$ matrices. This is illustrated 
explicitely in the example of this letter. 

\vskip2cm

Anomalies are undoubtedly one of the most important and intriguing aspects 
of Quantum Field Theory. Their existence is well established through theorems 
which assert them a topological origin \cite{1a}. Also, potentially anomalous 
contributions to a given theory can be pointed out by renormalization algebraic
techniques \cite{2a}. There remains however their explicit evaluation 
as a separate 
task. In the scrope of renormalizable models there are many ways to calculate 
anomalies: for example by regularizing the divergent amplitude with the 
constraints imposed by the renormalization prescription \cite{pok}. 
In a pioneer 
work Gernstein and Jackiw give an analytical evaluation of several Ward 
Identities \cite{geja}. 
In this work they 
resort to ambiguities inherent to performing shifts in linearly divergent 
integrals to explain the fact that it is not possible to simultaneously satisfy
all three Ward identities related to the anomalous pion decay \cite{geja} . 
This 
argument has ever since been quoted (and used) in many Field Thoery text books
on the subject. The essential difference between their approach and 
the one
in refs. \cite{pok} is that renormalization is not invoked 
as an essential 
ingredient 
for handling the divergent integrals. The physical meaning of the arbitrariness 
one has in choosing the internal momentum routing in loops is the translational 
invariance of free fermion propagators and therefore ascribing a fixed value 
to this arbitrary number ultimately means breaking one of the most fundamental 
symmetries of Q.F.T. In fact arbitrary momentum routings play no role in the  
evaluation of Ward identities when consistent regularizations are used, such as
Dimensional Regularization \cite{ref2}, proper-time \cite{ref2d}, 
Pauli-Villars \cite{ref1}, etc., since shifts 
in the integration variable are allowed. Unfortunately given the mathematical 
complexity of extending the definition of the $\gamma_5$ matrix to $\omega$ 
dimensions in Dimensional Regularization, an equally transparent discussion 
of the axial-vector-vector anomaly is not as easy \cite{ovrut}. Recently 
a technique 
has been proposed to manipulate divergent integrals directly in four dimensions
which makes use only {\bf implicitly} of a regulator and has proven to 
yield equivalent results as Dimensional Regularization with the advantage of 
being naturally applicable to amplitudes involving pseudo quantities. 

It is the purpose of the present contribution to calculate the 
axial-vector-vector amplitude in the context of such scheme. The technique 
has been proposed in ref. \cite{ori} and tested in several contexts 
\cite{PRD}. We 
sumarize the main steps of the prescription which we will follow here.

After evaluating Dirac traces wherever the case may be, one separates their 
divergent and finite contents. This separation is effected by means of 
implicitly assuming some regulating function from which two general properties
are essential: a) it should be even in momentum. b) a connection limit 
should exist. The presence of such a function is indicated in the integrals 
by the symbol $\Lambda$, but there is no need to specify it further. 
After assuming 
the presence of such a function use is made of mathematical identities, 
on the level of the integrand, to isolate divergent from finite contributions, 
(which contain the physics of the amplitude). The manipulations used are 
similar in spirit to those employed in BPHZ \cite{bphz}, with the 
essential difference 
that no subtraction is performed. The purpose here is to isolate 
the divergent content of the Feynman diagrams to be independent of external 
momenta  
and establish the set of basic divergent integrals 
for every theory. In this context renormalization is achieved without 
resorting to specific regularizations, but simply upon identifying the set 
of basic divergent objects, and proceeding for its reparametrization. The 
regularizing function is only implicitly assumed. 

Moreover, as extensively discussed in \cite{PRD},\cite{andre} 
there is a set of three 
Consistency Relations (CR) which should be satisfied in order to avoid 
ambiguities of any kind. They are 

\begin{eqnarray}
& &\int_{\wedge} \frac{d^4k}{(2\pi )^4}\frac{k_\mu k_\nu}
{(k^2-m^2)^2}=\frac{g_{\mu \nu}}{2}\int\frac{d^4k}{(2\pi )^4}
\frac{1}{(k^2-m^2)}=\frac{g_{\mu \nu}}{2}I_{quad}, \\
& &\int_{\wedge} \frac{d^4k}{(2\pi )^4}\frac{4k_\alpha k_\beta}
{(k^2-m^2)^3}=g_{\alpha \beta}\int\frac{d^4k}{(2\pi )^4}
\frac{1}{(k^2-m^2)^2}=g_{\alpha \beta}I_{log}, \\
& &\int_{\wedge} \frac{d^4k}{(2\pi )^4}\frac{k_\mu k_\nu k_\alpha k_\beta}
{(k^2-m^2)^4}=\frac{g_{\alpha \beta}}{2}\int\frac{d^4k}{(2\pi )^4}
\frac{k_\mu k_\nu }{(k^2-m^2)^3}.
\end{eqnarray}

\begin{eqnarray}
& &I_{quad}(m^2)=\int_{\wedge} \frac{d^4k}{(2\pi )^4}\frac{1}{(k^2-m^2)}, \\
& &I_{log}(m^2)=\int_{\wedge} \frac{d^4k}{(2\pi )^4}\frac{1}{(k^2-m^2)^2}, 
\end{eqnarray}

These relations must be satisfied by any regularization prescription which 
aims at consistency. Of course, they are satisfied by Dimensional 
Regularization, since they represent the 4-D equivalent of respecting 
translational invariance in perturbative expansions, mathematically 
materialized in the independence of arbitrary momentum routing in the 
internal loop lines. The existence of (at least) one such 4-D regulator has 
been proven in ref. \cite{andre} and therefore we do not discuss the validity 
of the Consistency Relations any longer. The obvious advantage of the 
present scheme is the fact that no special treatment needs to be given to the
$\gamma_5$ matrix.

In what follows we show how the anomalous pion decay comes about in 
4-D in a scheme where surface terms related to a specific choice 
of the momentum routing are absent by 
construction. A consistent treatment is given to divergent integrals and 
renormalizability is only invoked after all calculations have been performed.

We need and explicit evaluation of the following amplitude

\begin{eqnarray}
T^{AVV}_{\lambda \mu \nu} & = &  e^2 \int {d^4 k\over (2\pi)^4} 
Tr \{\gamma^{\lambda} \gamma_5 (\not k + \not p - m)^{-1} \gamma^{\mu} 
(\not k-m)^{-1} \gamma^{\nu} (\not k - \not q - m)^{-1}\} 
\end{eqnarray}

where the capital letters $AVV$ stand for Axial-Vector-Vector respectively and 
$\gamma_\mu$, are the usual Dirac matrices and $p,q$ the external momenta.
In order to obtain an analytical expression for the amplitude in eq. (4) 
and verify the relative Ward Identities with all external momenta off the mass
shell it is enough to evaluate the following three integrals. 
%
%
\begin{equation}
I_1 = \int_{\wedge} {d^4 k\over (2\pi)^4} \, {1\over (k^2 - m^2) 
[(p+k)^2 - m^2] \, [(q+k)^2 - m^2]} 
\label{31}
\end{equation}
\begin{equation}
I_2 = \int_{\wedge} {d^4 k\over (2\pi)^4} \, {k_{\mu}\over 
(k^2 - m^2) [(p+k)^2 - m^2] \, [(q+k)^2 - m^2]}
\label{29a}
\end{equation}
and
\begin{equation}
I_3 = \int_{\wedge} {d^4 k\over (2\pi)^4} \, {k_{\mu} k_{\nu}\over (k^2 - 
m^2) \, [(p+k)^2 - m^2] \, [(q+k)^2 - m^2]}
\label{30a}
\end{equation}

The first two are finite and can be evaluated essencially by any method 
to yield

\begin{equation}
I_1 = {i\over (4\pi)^2} \, \xi_{00}(p,q)  
\label{31a}
\end{equation}

\begin{equation}
I_2 = - {i\over (4\pi)^2} \{q_{\mu} \xi_{10}(p,q) + p_{\mu} \xi_{01}(p,q)\} 
\label{32}
\end{equation}

where the functions $\xi_{nm}$ are defined in the Appendix. 

The last integral eq. (7) is logarithmic divergent and here care must be 
exercised if one does not consider eventual substractions from the beginning. 
We give the steps for the evaluation of (7) according to our prescription: we
first use a mathematical identity at the level of the integrand in order 
to isolate the purely (external momenta independent) part of the integral, 

and
\begin{equation}
I_3 = I^{div}_{\mu \nu} + I^{fin}_{\mu \nu} 
\label{33}
\end{equation}
where
\begin{equation}
I^{div}_{\mu \nu} = {g_{\mu\nu}\over 4} \int_{\wedge} {d^4 
k\over (2\pi)^4} {1\over (k^2 - m^2)^2}
\label{33a}
\end{equation}

\begin{eqnarray}
I^{fin}_{\mu \nu} = & - & {i\over (4\pi)^2} \{g_{\mu \nu} 
[{1\over 2} Z_0 ((p-q)^2; m^2)\nonumber\\ & - & ({1\over 2} + 
m^2 \xi_{00}(p,q)) + {q^2\over 2} \xi_{10}(p,q) + {p^2\over 2} 
\xi_{01}(p,q)]\nonumber\\ & - & p_{\mu} p_{\nu} \xi_{02}(p,q) - q_{\mu}
q_{\nu} \, \xi_{20}(p,q)\nonumber\\ & - & (p_{\mu} q_{\nu}  - 
p_{\nu} q_{\mu}) \xi_{11}(p,q)\}
\label{33b}
\end{eqnarray}
where the functions $Z_k$ and $\xi_{nm}$ depend on the external momenta
$p,q$ and on the mass $m$, see Appendix. 
\vskip0.5cm

In order to verify the Ward identities we need to observe the following 
relations between the functions $\xi_{nm}$.
\begin{equation}
p^2 \xi_{10}(p,q) - p.q \xi_{01}(p,q) = {1\over 2} \{Z_0 (q^2; m^2) - 
Z_0 (p.q; m^2) + p^2 \xi_{00}(p,q)\}
\label{40a}
\end{equation}
\begin{equation}
q^2\xi_{11}(p,q) + p.q\xi_{02}(p,q) = {1\over 2} \left\{{-Z_0(p - q)^2;m^2)
\over 2}
+ {Z_0(p^2; m^2)\over 2} + q^2 \xi_{01}(p,q)\right\}
\end{equation}
\begin{equation}
q^2\xi_{20}(p,q) + p.q\xi_{11}(p,q) = {1\over 2}\left\{-\left[{1\over 2} + m^2 
\xi_{00}(p,q)\right] + {p^2\over 2} \xi_{01} (p,q) + {3q^2\over 2} 
\xi_{10}(p,q)\right\}
\end{equation}
\begin{equation}
p^2\xi_{02}(p,q) + p.q\xi_{11}(p,q) = {1\over 2} \left\{-\left[{1\over 2} 
+ m^2 \xi_{00}(p,q)\right] + {q^2\over 2} \xi_{10}(p,q) + 
{3p^2\over 2} \xi_{01}(p,q)\right\}
\end{equation}
\begin{equation}
p^2\xi_{11}(p,q) + p.q\xi_{20}(p,q) = {1\over 2} \left\{-{1\over 2} 
Z_0((p-q)^2;m^2) + {1\over 2} Z_0(q^2;m^2) + p^2 \xi_{10}(p,q)\right\}
\end{equation}
\vskip0.5cm

After long and tedious but straightforward algebraic procedure we get 

\begin{eqnarray}
T^{AVV}_{\lambda \mu \nu} & = &  e^2 \int {d^4 k\over (2\pi)^4} 
Tr \{\gamma^{\lambda} \gamma_5 (\not k + \not p - m)^{-1} \gamma^{\mu} 
(\not k-m)^{-1} \gamma^{\nu} (\not k - \not q - m)^{-1}\}\nonumber\\ & = & 
e^2\{\epsilon_{\lambda \mu \nu \omega} (p_{\omega} - q_{\omega}) F_1 (p,q)
\nonumber\\ & + & p_{\omega} q_{\phi} \{[\epsilon_{\lambda \nu \omega \phi} 
q^{\mu} + \epsilon_{\lambda \mu \omega \phi} q^{\nu}]F_2 (p,q) \nonumber\\
& + & [\epsilon_{\lambda\nu \omega \phi} p^{\mu} + 
\epsilon_{\lambda \mu \omega \phi} p^{\nu}] F_3 (p,q)\nonumber\\
& + & \epsilon_{\mu \nu \omega \phi} (p^{\lambda} F_4 (p,q) + q^{\lambda}
F_5 (p,q)\}\nonumber\\ & + & \epsilon_{\lambda \mu \nu \omega} p_{\omega} 
F_6 (p,q) - \epsilon_{\lambda \mu \nu \omega} q_{\omega} F_7 (p,q)\}
\end{eqnarray}
where $A$ stands for axial vector and
\begin{equation}
F_1 (p,q) = -{1\over 4\pi^2} \left[{Z_0\over 4} ((p+q)^2; m^2) - 
{1\over 4} + {m^2 \xi_{00}(p,q)\over 2} + {q^2 \xi_{01}(p,q) + 
p^2 \xi_{10}(p,q)\over 
4}\right] 
\label{37}
\end{equation}
\begin{equation}
F_2 (p,q) = {1\over 4\pi^2} [\xi_{01}(p,q) - \xi_{02}(p,q) - \xi_{11}(p,q)] 
\label{37b}
\end{equation}
\begin{equation}
F_3 (p,q) = - {1\over 4\pi^2} [\xi_{11}(p,q) + \xi_{20}(p,q) - \xi_{10}(p,q)]
\label{37c}
\end{equation}
\begin{equation}
F_4 (p,q) = - {1\over 4 \pi^2} [\xi_{11}(p,q) + \xi_{10}(p,q) - \xi_{20}(p,q)]
\label{37d}
\end{equation}
\begin{equation}
F_5 (p,q) = - {1\over 4\pi^2} [\xi_{11}(p,q) + \xi_{01}(p,q) - \xi_{02}(p,q)]
\label{37e}
\end{equation}
\begin{equation}
F_6 (p,q) = - {1\over 4\pi^2}\left[-{Z_0(p^2; m^2)\over 4} - {(p+q)^2
\over 2} \xi_{10}(p,q)\right]
\label{37f}
\end{equation}
\begin{equation}
F_7 (p,q) = - {1\over 4\pi^2} \left[{Z_0 (q^2; m^2)\over 4} - {(p+q)^2
\over 2} \xi_{01}(p,q)\right]
\label{37g}
\end{equation}

Now, in the context of a renormalizable theory such as the Linear Sigma 
Model there is one step still missing: the divergent contribution
$I_{log}$ must be renormalized away while a finite part must be fixed in a 
{\bf unique} way \cite{geja}. At this point the essential meaning 
of the word {anomaly} 
appears very clearly: it is impossible to renormalize this amplitude 
simultaneously preserving all three Ward-Identities.

Note that in the present prescription the evaluation of eq. (0.20) which 
contains
a bilinear in the numerator could not be performed in a different way, e.g., 
by considering contractions of the indices, $\mu, \nu$  since here this 
procedure is not consistent 
with the CR eqs. (0.1-0.3). The difference between the two 
calculations is a constant and therefore it is of no importance in the context 
of renormalizable theories.

In summary we have calculated the triangle axial-vector-vector amplitude 
with all external momenta off shell which is analytic up to one scalar 
integral $\xi_{00}$, related to the Spence function. We have shown that there 
exist relations between the functions used to systematize the results which are
essential to verify the Ward identities. Our calculation is performed within 
a 4-D scheme where an eventual regulator needs never be explicitated and  
odd numbers of $\gamma_5$ matrices are easy to handly.

\vskip0.5cm

{\bf Appendix}
\vskip0.5cm

{\bf General Integrals for the Finite Content of One Loop 
Amplitudes}

\vskip0.5cm

{\bf The Functions $Z_k (p^2; m^2)$}
\vskip0.5cm

We define 
\begin{equation}
Z_k (p^2; m^2) = \int^1_0 dz \, z^k ln \, \left({p^2 z (1-z)-m^2
\over - m^2}\right)
\label{1}
\end{equation}
where $k$ is an integer, $m$ is the mass parameter which appears in the 
propagators, $p$ is some external momentum.  %

\vskip0.5cm

{\bf The Functions $\xi_{n m}:$}

\vskip1cm

When we consider one loop Feynman integrals associated to three point 
functions with two external momenta and three masses, the finite parts of 
the amplitudes are always related to the following general structures
\begin{equation}
\xi_{n m}(p,q) = \int^1_0 \, dz \int^{1-z}_0 
\, dy {z^n y^m\over Q (y, z)}
\label{22}
\end{equation}
where
\begin{equation}
Q = (y, z) = p^2 y (1-y) + y + q^2 z (1-z) + - m^2_1 - 2 \, p.q\,  y z
\end{equation}
In what follows we present the analytical expressions for equal masses and 
the $k =0,1,2$ cases, which will appear in our calculations. 
An extensive account of such structures in more general situations 
is given in reference \cite{ori}. 
All the $\xi_{n m}$ functions we need will turn out to become linear 
combinations of the $Z_k$ functions defined previously and the function
$\xi_{00}$, related to the Spence function,
\begin{equation}
{i\over (4\pi)^2} \xi_{00}(p,q) = {Z_{-1} ((p-q)^2 ;m^2 )\over (p-q)^2} =
\int^1_0 \frac{dz}{z} ln (\frac{(p-q)^2 z(1-z)-m^2}{-m^2}) 
\label{23}
\end{equation}
The procedure is the same as that of the preceeding section. We have 
\begin{eqnarray} 
\xi_{01}(p,q) & = &  \{{p^2 
q^2\over p^2 q^2 - (p.q)^2}\} \{{(p^2 - p.q)\over 2 p^2 
.q^2} [(-) Z_0 (m^2, (p-q^2); m^2)] 
\nonumber\\& + & {1\over 2 p^2} [Z_0 (p^2; 
m^2)]\nonumber\\ & + & {(p.q)\over 2p^2.q^2} [(-) Z_0 (q^2; m^2)]
\nonumber\\ & + & {(p^2 - p.q)\over 2 p^2} [\xi_{00}(p,q)]\}  
\label{24}
\end{eqnarray}
\begin{eqnarray}
\xi_{10}(p,q) & = & \xi_{01}(q,p)\   
\label{25}
\end{eqnarray}
\begin{eqnarray}
\xi_{02}(p,q) & = & \{{p^2 
q^2\over p^2 q^2 - (p.q)^2}\} \{{(p.q)\over 4p^2 q^2} 
[Z_0 ((p-q)^2; m^2) -  
[Z_0 (p^2; 
m^2)]\nonumber\\ & + & {(p^2-p.q)\over 2p^2} [\xi_{01}(p,q)] - {1\over 2p^2} 
\left[{1\over 2} + m^2 \xi_{00}(p,q)\right] \nonumber\\ & + & {1\over 4 p^2} 
[q^2 (\xi_{10}(p,q)) + p^2 (\xi_{01}(p,q))]\}
\label{27}
\end{eqnarray}

\begin{eqnarray}
\xi_{20}(p,q) & = & \xi_{02}(q,p)\   
\label{25}
\end{eqnarray}
\begin{eqnarray}
\xi_{11}(p,q) & = & \{{p^2 q^2\over p^2 q^2 - (p.q)^2}\} \{(-) {1\over 4q^2} 
[Z_0 ((p-q)^2; m^2) - Z_0 (p^2; m^2)] \nonumber\\
& + & {(q^2-p.q)\over 2q^2} [\xi_{01}(p,q)] + {(p.q)\over 2p^2.q^2} 
[{1\over 2} + m^2 \xi_{00}(p,q)]\nonumber\\ & + & -{(p.q)\over 4 p^2 q^2} 
[q^2 (\xi_{10}(p,q)) + p^2 (\xi_{01}(p,q))]\} 
\label{28}
\end{eqnarray}

\vskip0.5cm

{\bf Acknowledgments}: This work has been partly supported by CNPq, 
FAPEMIG, Funda\c{c}\~{a}o para a Ci\^encia e a Tecnologia FCT,
PRAXIS XXI, BCC/4301/94, PCERN/FIS/1086/96 and /1162/97, PESO/S/PRO/
1057/95

\end{document}